**PyGRF: An improved Python Geographical Random Forest model and case studies in public health and natural disasters**

Kai Sun [a], Ryan Zhenqi Zhou [a], Jiyeon Kim [a], and Yingjie Hu [a, b]

[a.] *GeoAI Lab, Department of Geography, University at Buffalo, Buffalo, NY, USA*
[b.] *Department of Computer Science and Engineering, University at Buffalo, Buffalo, NY, USA*

**Abstract**: Geographical random forest (GRF) is a recently developed and spatially explicit machine learning model. With the ability to provide more accurate predictions and local interpretations, GRF has already been used in many studies. The current GRF model, however, has limitations in its determination of the local model weight and bandwidth hyperparameters, potentially insufficient numbers of local training samples, and sometimes high local prediction errors. Also, implemented as an R package, GRF currently does not have a Python version which limits its adoption among machine learning practitioners who prefer Python. This work addresses these limitations by introducing theory-informed hyperparameter determination, local training sample expansion, and spatially-weighted local prediction. We also develop a Python-based GRF model and package, PyGRF, to facilitate the use of the model. We evaluate the performance of PyGRF on an example dataset and further demonstrate its use in two case studies in public health and natural disasters.

**1 Introduction**

Geographical random forest (GRF) is a spatially explicit machine learning model that has been developed recently (Georganos et al., 2021; Georganos & Kalogirou, 2022). As a spatial extension of the general Random Forest (RF) model, GRF borrows the idea of geographically weighted regression (GWR) (Brunsdon et al., 1998; Fotheringham et al., 2003) by fitting a local RF model at the location of each data instance using the nearby data within a specified bandwidth. In addition to the local RF models, GRF also fits a global RF model using the entire dataset during the training phase. To make a prediction, GRF combines the prediction from the local RF model that is closest to the test data and the prediction from the global RF model using a weighted approach.

GRF has two main advantages compared with the typical and non-spatial RF model. First, it can improve the prediction accuracy of the non-spatial RF model. GRF fuses the prediction of the global RF model which has learned the general data pattern from the entire dataset and the prediction from the local RF model which has learned the local specific pattern. As a result, GRF can often make more accurate predictions compared with the non-spatial RF model, although the improvement can be small (Georganos et al., 2021). Second, GRF enables the exploration of local feature importances and their variations across geographic space. Unlike the non-spatial RF model that derives only a single set of global feature importance from the data, GRF fits many local RF models across the study area which reveal local feature importance at different locations and how they vary spatially. Given these two advantages, GRF has already been used in a wide range of studies, such as predicting population density (Georganos et al., 2021), identifying factors related to COVID-19 death (Grekousis et al., 2022), examining leaf functional traits (Aguirre-Gutiérrez et al., 2021), and assessing landslide susceptibility (Quevedo et al., 2022).

While GRF is useful, our previous research has identified three limitations of the current model (Zhou et al., 2022). First, the optimal values of two important hyperparameters of GRF, i.e., the bandwidth and local model weight, are determined in a trial-and-error fashion which is computationally-intensive and time-consuming. Second, given the reduced number of local training data instances at each location, a local RF model may not be sufficiently trained and may output inaccurate local predictions. Third, the current GRF model uses only the one single closest local RF model to provide the local prediction, and such a local prediction could have low accuracy depending on the quality of that one particular local RF model. In addition, implemented as an R package, GRF currently does not have a Python version which limits its adoption among many machine learning practitioners who prefer to use Python. A recent work by Wiedemann et al. (2023) provided a preliminary Python implementation of GRF as a byproduct of their research. While providing a nice first-step implementation, Wiedemann et al. (2023) did not implement some important features of the model, such as spatial weighting of the training data, and did not conduct evaluations to ensure that the Python implementation produces results consistent with the results of the original R-based GRF model.

Building on the existing GRF research (Georganos et al., 2021; Georganos & Kalogirou, 2022; Wiedemann et al., 2023) and our previous work (Zhou et al., 2022), we develop PyGRF as a Python-based GRF model. We implement this Python model based on a careful study of the source code of the original model in R, and further conduct experiments to ensure consistency between the outputs of our Python implementation and the R model. More importantly, we address the limitations of the current GRF model by introducing three model improvements. First, we propose a theory-informed hyperparameter determination approach to help determine the bandwidth and local weight hyperparameter values by assessing the spatial autocorrelation of the data, which can substantially reduce the time needed for hyperparameter tuning. Second, we provide a local training sample expansion strategy based on bootstrapping to increase the size of local training data and better fit local models. Third, we develop a spatially-weighted local prediction approach to reduce prediction errors due to one single local RF model by combining the predictions from multiple local RF models using spatial weights. The contributions of this paper are as follows:

- We propose three model improvements to address the limitations of GRF and evaluate their effectiveness based on an example dataset. These improvements are incorporated in our newly developed Python version of the GRF model, PyGRF.

- We conduct two case studies in public health and natural disasters using the PyGRF model and package. The case studies demonstrate the use of PyGRF in real-world domain problems and provide further evaluations on the proposed model improvements.

- We publish our implemented PyGRF via the Python package system *pip* at: https://pypi.org/project/PyGRF. We also share the source code of PyGRF and the two case studies on GitHub at: https://github.com/geoai-lab/PyGRF.

The remainder of this paper is organized as follows. Section 2 provides a review of the background and research related to spatially explicit models and GRF. Section 3 presents the methodological details of our proposed three model improvements. In Section 4, we evaluate the consistency between our PyGRF and the original R-based GRF model, and we also evaluate the effectiveness of the model improvements. In Section 5, we conduct two case studies by using the developed PyGRF model to estimate neighborhood-level obesity prevalence and to predict help requests to prepare for future winter storms. Section 6 discusses the effectiveness of the proposed model improvements, and finally, Section 7 concludes this work.

**2 Related work**

Spatially explicit models have received much attention from the GIScience community. Goodchild (2001) proposed four tests to examine whether a model can be considered as spatially explicit, which are: (i) whether the model result varies across different locations (*variance test*); (ii) whether the model contains spatial representations (*representation test*); (iii) whether the model uses spatial concepts in its formulation (*formulation test*); and (iv) whether the model output has a different spatial structure compared with the input (*outcome test*). A model that passes one or multiple of the tests can be considered as a spatially explicit model. By capturing the underlying spatial process and local patterns in geographic data, spatially explicit models have been developed and used in many applications from public health to ecology (DeAngelis & Yurek, 2017; L. Li, 2019; O'Sullivan et al., 2020).

A number of spatially explicit statistical models have been developed in the literature. Examples include the spatial lag and spatial error regression models (Anselin, 2009), eigenvector spatial filtering (Griffith, 2003), GWR (Brunsdon et al., 1998; Fotheringham et al., 2003), and the more recent multi-scale GWR (Fotheringham et al., 2017). These statistical models have played highly important roles in geographical analysis research by accommodating spatial autocorrelation and spatial heterogeneity commonly existing in geographic data, and can provide more robust analysis results than traditional non-spatial ordinary least squares (OLS) regression. Spatial regression models, however, cannot effectively model non-linear relationships between independent and dependent variables (Wiedemann et al., 2023). Nevertheless, spatial regression models are still widely used in many studies, thanks to their transparent model architecture and high model explainability.

With the fast advancement of geospatial artificial intelligence (GeoAI), researchers have also developed spatially explicit machine learning models (Janowicz et al., 2020; Mai et al., 2022; Hu et al., 2024). Examples include place2vec (Yan et al., 2017), geographically weighted artificial neural network (Hagenauer & Helbich, 2022), geographically and temporally weighted neural network (Feng et al., 2021), GRF (Georganos et al., 2021), and many others (Gupta et al., 2021; Islam et al., 2021; Masrur et al., 2022). One advantage of machine learning models over typical linear regression models is their ability to handle non-linear relationships, which often result in higher prediction accuracy. A main disadvantage is their limited model explainability, although explainable AI frameworks, such as Shapley Additive Explanations (SHAP), have been increasingly used to improve model explainability (Z. Li, 2022).

Random forest is a nonparametric ensemble machine learning model (Ho, 1995). It trains a group of decision trees with randomness and makes final predictions by combining the predictions from individual decision trees. Random forest adds randomness into the construction of individual trees by training each tree with a random sample of the training data and using only a random subset of features at each node of a tree. As a result, RF is less likely to overfit compared with a single decision tree model, due to the introduced randomness and the combination of predictions from multiple trees (Breiman, 2001). While deep learning models have received much attention in recent years, existing research has shown that RF models often provide more accurate predictions on structured tabular data (i.e., data formatted as rows and columns in a comma-separated values file) than deep learning models (Gao et al., 2019; Hu et al., 2021; Chang et al., 2022), although deep learning models usually perform better on imagery and textual data. In addition to its high prediction accuracy on structured tabular data, RF also outputs feature importance for explaining the usefulness of input features for making predictions. Accordingly, RF offers higher model explainability than a typical deep neural network model.

Geographical random forest extends the RF model by training a local RF model at the location of each data instance using nearby data within a distance. The distance value is defined via a bandwidth hyperparameter $\lambda$. In addition to the local models, GRF also trains a global RF model using the entire dataset, and the prediction of GRF is based on a weighted combination of the predictions from both the global model and the closest local model using Equation (1):

$$y_i = \alpha * y_{li} + (1 - \alpha) * y_{gi} \qquad (1)$$

where $y_i$ is the final prediction of the GRF model for the $ith$ data instance in the test data, $y_{li}$ is the prediction of the closest local RF model, and $y_{gi}$ is the prediction of the global RF model. $\alpha$ is the local weight hyperparameter whose value range is in [0, 1]. A higher $\alpha$ value puts more weight on the local model, while a lower $\alpha$ value puts more weight on the global model. The GRF model becomes a completely local model when $\alpha=1$, and it can also become a completely global model (i.e., a regular RF) when $\alpha=0$.

As a spatial extension of the RF model, GRF inherits the merit of RF in good prediction accuracy and model explainability, while extending the global feature importance of RF to local importance across different locations. GRF passes all four tests for spatially explicit models: (i) its predictions vary across different locations, (ii) it represents distance decay in the training data via spatial weighting, (iii) it uses the concept of neighborhood to train local models, and (iv) its output often has a different spatial structure compared with that of the input data. While useful, the current GRF model has limitations in determining hyperparameters, training local models, and making accurate local predictions. This work, therefore, addresses these limitations by proposing three model improvements. We also develop a Python version of the GRF model, called PyGRF, to incorporate these model improvements.

## 3 Methods

In this section, we discuss the limitations identified from the current GRF model, and present corresponding model improvements to address these limitations.

### 3.1 Theory-informed hyperparameter determination

The performance of GRF is sensitive to the values of two hyperparameters, bandwidth $\lambda$ and local weight $\alpha$. Bandwidth $\lambda$ affects the data instances used to train local models (i.e., only those data instances within the bandwidth distance are used to train the local models), and local weight $\alpha$ affects the relative weights put on the local model and the global model when they are combined to make predictions. The original research of GRF has shown that improper values of these two hyperparameters can lead to inferior performance of the model (Georganos et al., 2021). To determine suitable values for the two hyperparameters, the authors of GRF proposed a trial-and-error hyperparameter tuning approach that iteratively tries different bandwidth values from the 0.05 quantile of the total number of data instances to the 0.95 quantile and also iteratively tries three discrete values of 0.25, 0.5, and 0.75 for the local weight $\alpha$ (Georganos et al., 2021; Georganos & Kalogirou, 2022). While such an approach can identify good values for the two hyperparameters, it is computationally intensive and time consuming based on our experience of using the model. The GRF model itself already has a high computational cost since it fits many local RF models across different locations, and this trial-and-error hyperparameter tuning process further increases the computational cost of using the model.

To address this limitation, we propose a theory-informed approach based on spatial autocorrelation to determine bandwidth $\lambda$ and local weight $\alpha$. Instead of directly training and testing many GRF models based on different hyperparameter values, we propose to first understand the spatial autocorrelation in the data and the spatial scale at which spatial autocorrelation is most significant. The rationale is that the spatial scale with the most significant spatial autocorrelation can suggest a suitable bandwidth distance $\lambda$ within which local data instances are most similar and can be used for training effective local models; meanwhile, the extent of spatial autocorrelation can suggest a suitable local weight $\alpha$, since a stronger spatial autocorrelation indicates a higher similarity among local data instances which will likely contribute to more effective local models. We utilize the technique of incremental spatial autocorrelation which measures the global Moran's I index and the associated z-score based on a sequence of incrementally increasing distances. The global Moran's I index provides an overall score in the value range of [-1, 1] to quantify the spatial autocorrelation of the data at a given distance, while the z-score indicates the significance of such autocorrelation. The distance at which the z-score is the highest is used as the value for bandwidth $\lambda$, and the global Moran's I index at that distance is used as the value of local weight $\alpha$, based on Equation (2):

$$\alpha = \begin{cases} Moran's\ I, & if\ Moran's\ I > 0\ and\ p < 0.05 \\ 0, & otherwise \end{cases} \quad (2)$$

As shown in the equation, the local weight $\alpha$ is set to the value of global Moran's I when it is statistically significant and larger than 0. In these situations, nearby data points in a local area share similar values, and the local RF models are more likely to capture local patterns and output more accurate local predictions. Further, a higher Moran's I index will give a higher weight to the local RF models to better utilize the captured local patterns. Note that a positive and significant global Moran's I index is in the value range of (0,1], which matches the value range of local weight $\alpha$ needed for the GRF model. In situations when the global Moran's I is not statistically significant or is negative, nearby data points in a local area do not share similar values or are more or less randomly distributed. Consequently, the local RF models are unlikely to be effective since there do not exist clear local patterns. In those situations, our approach will set the local weight $\alpha$ to 0, which turns the GRF model into a regular RF model that utilizes all information of the entire dataset to make predictions. This theory-informed approach has a considerably lower computational cost than the current trial-and-error hyperparameter tuning approach. This is because our approach only assesses the spatial autocorrelation of the data, and does not train a large number of GRF models for different possible hyperparameter settings as done in the trial-and-error hyperparameter tuning approach.

### 3.2 Local training sample expansion

While the local RF models of GRF are designed to capture local patterns of the data, they can run into the difficulty of insufficient local training samples. Since only the data instances within the bandwidth are used to train a local model, the size of the local training data can be very small, e.g., 10 data instances, depending on the bandwidth hyperparameter set by the model user. This small size of local training data can be insufficient for training a local RF model, especially when the model is fairly complex. For example, 10 data instances at a local location are unlikely to be sufficient for training an RF model that has 100 decision trees. These insufficiently trained local RF models can lead to inaccurate local predictions which further decrease the performance of the whole GRF model.

To mitigate this issue, we propose a local training sample expansion strategy to increase the size of local training data. Since we aim to create a larger local training dataset, we use bootstrapping which is a commonly used resampling method that repeatedly samples data from the original dataset with replacement. This method allows us to create a larger simulated training dataset while ensuring that each added data instance is real. To reduce the risk of overfitting, we limit the size of the expanded dataset to be either two times the size of the original local data or two times the number of trees in the local RF model, depending on which number is smaller. Limiting the size of the expanded dataset also reduces the extra computational cost related to data resampling and model training on larger local datasets. Further, there also exist situations when a large bandwidth is specified and the size of local training data is likely to be sufficient for training the local RF model. In such situations, we directly use the original local data for model training without performing further data expansion. We formalize this local training sample expansion strategy as Equation (3):

$$Expanded\ D = \begin{cases} Bootstrap(D, size = \min(2*ntree, 2*|D|)), & |D| < 2*ntree \\ D, & otherwise \end{cases} \quad (3)$$

where $D$ represents the original local dataset (which contains the data instances within the specified bandwidth distance). If the size of $D$ is smaller than two times the number of trees, the local dataset will be expanded by bootstrapping it to the size of the smaller number of 2*$ntree$ (two times the number of trees of the RF model) or 2*$|D|$; otherwise, the original local dataset will be directly used to train the local RF model. We note that the proposed local training sample expansion strategy does not completely eliminate the issue of insufficient training data when the bandwidth is small, since the resampled data do not introduce new information to the model. However, this strategy can help mitigate this issue.

## 3.3 Spatially-weighted local prediction

The current GRF model uses only one single closest local model to make the local prediction. This local prediction is then combined with the global prediction to become the final prediction of the GRF model. Depending on the specific local data instances used to train this one local RF model, its prediction could have large errors. For example, if a data outlier exists in the local region, the trained local RF model can be largely affected by this one data outlier. Meanwhile, other local RF models that are fairly close to the target location (e.g., the local RF models that are the 2nd and 3rd closest to the target location) are likely to provide useful information for prediction as well, but they are not utilized in the current GRF model.

Based on these considerations, we propose spatially-weighted local prediction that uses an ensemble approach to compute the local prediction. Instead of using only the one closest local RF model, we use all local RF models within the bandwidth and combine their predictions in a spatially weighted manner. The rationale of using spatial weights to combine local model predictions is that nearby local models are more likely to provide useful predictions than distant models, following the First Law of Geography (Tobler, 1970). We formalize this approach using Equation (4):

$$y_{li} = \frac{\sum_{j=1}^{j \leq \lambda} w_{ij} * y_{ij}}{\sum_{j=1}^{j \leq \lambda} w_{ij}} \quad (4)$$

where $y_{li}$ is the local prediction for the $ith$ data instance in the test dataset, $y_{ij}$ is the prediction from the $jth$ nearby local RF model for the $ith$ data, and $w_{ij}$ is the spatial weight determined by the distance between locations $i$ and $j$. We use the same kernel as used in the original GRF model for model training (i.e., the bisquare kernel) to compute $w_{ij}$, and the predictions from closer local RF models are assigned higher weights than those from farther away local RF models. Since this spatially-weighted local prediction combines the predictions from all local RF models within the bandwidth, it is less susceptible to data outliers affecting one particular local RF model. We note that a similar idea of spatially-weighted prediction was also proposed in Wiedemann et al. (2023). Their approach uses local decision tree models whose locations are determined based on a K-means clustering process, while our approach uses local RF models fitted at each data instance within the bandwidth.

## 4 Implementation and evaluation experiments

### 4.1 Implementation

We implement the PyGRF model and package based on a careful study of the source code of the original R-based model implemented by the GRF authors (Georganos & Kalogirou, 2022). The main Python libraries used in our implementation include *scikit-learn* (for implementing the RF model) and *pysal* (for implementing spatial operations such as computing spatial autocorrelation). We implement all features of the GRF model, including spatial weighting and the more advanced parallel computing feature. We also provide flexibility in our implementation by allowing the user to turn on and off any of the three proposed model improvements. Thus, a user can choose to simply use the original GRF model in Python that is consistent with the R-based GRF model, or the user can choose to turn on any or all of the three proposed model improvements. We further conduct unit tests on our code to ensure its stability and robustness. All functions and modules have passed the tests. The test results are provided in Supplementary Figure S1. We publish PyGRF via the Python package management system *pip* at: https://pypi.org/project/PyGRF, and an interested reader can quickly install this package via "pip install PyGRF". We also share the source code of PyGRF and a detailed description of its functions and parameters on GitHub at: https://github.com/geoai-lab/PyGRF. Two Jupyter Notebooks for the two case studies in Section 5 are also shared in this GitHub repository, which serve as tutorials for using the package.

### 4.2 Evaluation of consistency between PyGRF and the R-based GRF

While we have implemented PyGRF by following the source code of the original R package, our implementation is nevertheless based on a different programming language and uses a set of different Python-based libraries (e.g., *scikit-learn* and *pysal*). To ensure that PyGRF can perform in a consistent manner as the original R-based package, we conduct experiments to compare the outputs of the two implementations using the example dataset provided in the R package of GRF. This example dataset contains mean household income at the municipal level in Greece in 2011, and three independent variables are included to predict the mean household income, which are: unemployment rate, primary sector employment proportion, and non-Greek citizens proportion. This dataset has 325 records in total.

To compare the two implementations, we train PyGRF and the R-based GRF using the exact same hyperparameter setting: we set the local weight $\alpha$ as 0.5 to combine the local and global predictions in a balanced manner; we set the bandwidth $\lambda$ as 60 which is tuned using the original R-based GRF with 70% of the data. To add some variations in our experiments, we compare the two implementations using three different numbers of trees for the RF models, which are 50, 75, and 100 trees respectively. Due to the randomness in RF, a random seed is needed for constructing the model. However, since Python and R use different randomness generation mechanisms, the same random seed can still generate different results in the two implementations. To increase the robustness of our experiments, we generate 100 random seeds based on a uniform distribution, and run our experiments 100 times using different random seeds. The final prediction for each test data instance is obtained by averaging the predictions from the 100 experiments. We then compare the final predictions from PyGRF and the R-based GRF model to assess their consistency.

Figure 1 shows the predictions and prediction errors from the two implementations with the three different numbers of trees. As can be seen in subfigures (a), (b), and (c), the predictions of the two implementations are highly consistent and the points in the scatter plots are all located close to the reference line in the diagonal. Similar results are observed in the prediction errors of the two model implementations in subfigures (d), (e), and (f). We further quantify this consistency using Pearson's correlation. The correlation coefficients ($\rho$) of the predictions and prediction errors across the three settings are all close to 1. These results indicate a strong consistency between the outputs of the two implementations. Note that there exist tiny differences in Pearson's correlation coefficients in the three settings, but the differences are beyond the third digit and thus the same $\rho$ values are shown on the scatter plots.

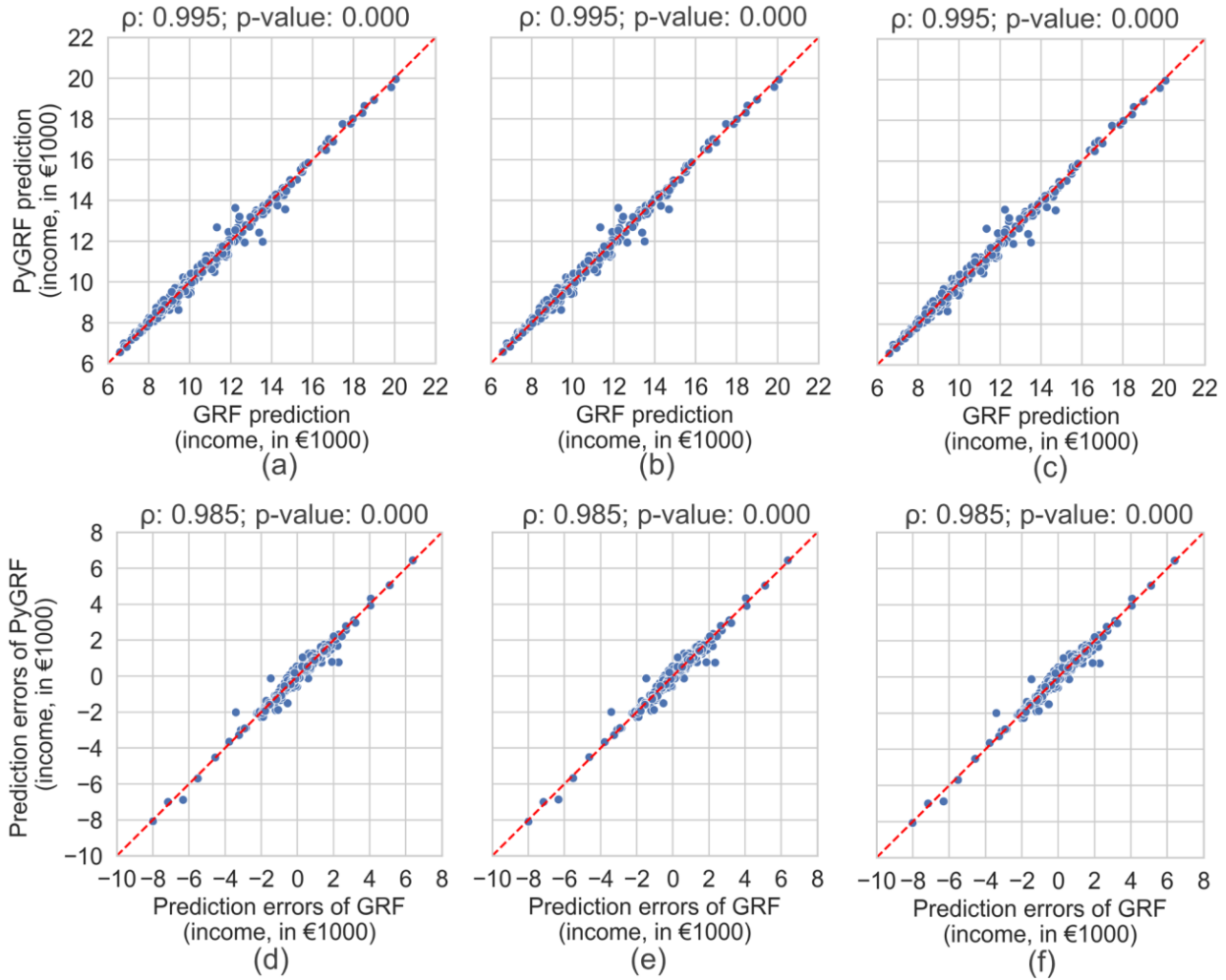

Figure 1. Consistency evaluation between PyGRF and the R-based GRF: (a), (b) and (c) are predictions of the two model implementations with 50 trees, 75 trees, and 100 trees respectively; (d), (e), and (f) are prediction errors of the two model implementations with 50 trees, 75 trees, and 100 trees.

### 4.3 Evaluation of the model improvements

We continue to evaluate the effectiveness of the proposed model improvements. Since our previous experiments have shown that the outputs of PyGRF are consistent with those of the R-based GRF model, the following experiments use PyGRF directly. We still use the example income dataset in these experiments, and three sets of experiments are conducted:

- ***Comparison with the RF model*:** in this set of experiments, we compare the PyGRF model with the RF model to demonstrate the performance improvement brought by PyGRF without adding any model improvement.
- ***Effectiveness of each improvement*:** in this set of experiments, we evaluate the effectiveness of each proposed model improvement individually by assessing the performance of PyGRF with and without a model improvement.

- ***Performance with each improvement added step-by-step***: in this set of experiments, we evaluate the effectiveness of the proposed model improvements incrementally by adding each improvement step-by-step and examining model performance changes.

Two metrics are used for assessing model performance: R-squared ($R^2$) and root mean square error (*RMSE*). The two metrics are defined in the Equations (5) and (6), where $y_i$ is the true value of the *ith* data instance, $\hat{y}_i$ is the prediction of the model, and $\bar{y}$ is the mean of true values.

$$R^2 = 1 - \frac{\sum_i(y_i - \hat{y}_i)^2}{\sum_i(y_i - \bar{y})^2} \tag{5}$$

$$RMSE = \sqrt{\frac{\sum_{i=1}^{n}(y_i - \hat{y}_i)^2}{n}} \tag{6}$$

Four hyperparameters need to be set in order to run the models, which are: *ntree* (the number of trees for the RF model), *mtry* (the number of maximum features to be tried at each split of a decision tree), $\lambda$ (bandwidth), and $\alpha$ (local model weight). The former two hyperparameters are from the RF model, while the latter two hyperparameters are introduced by the GRF model. For the default PyGRF model, we use the current trial-and-error hyperparameter tuning approach via grid search to identify suitable values for the four hyperparameters. The search spaces are defined as follows: *ntree*: (0, *N/2*] with an interval of 20, where *N* is the number of samples in the training data; *mtry*: {1, 2, 3}, considering that there are only three independent variables in this example data; $\lambda$: [0.05 quantile, 0.95 quantile] of samples with an interval of 5; $\alpha$: {0.25, 0.5, 0.75}. These search spaces are defined following the original paper of GRF and its source code (Georganos et al., 2021; Georganos & Kalogirou, 2022).

While the default PyGRF model needs to tune all four hyperparameters (consistent with the original R-based model), our proposed theory-informed hyperparameter determination can help choose values for the bandwidth and local weight parameters via incremental spatial autocorrelation. Figure 2 shows the incremental spatial autocorrelation result based on the example income dataset. As can be seen, the z-score of spatial autocorrelation achieves the highest value when the bandwidth equals 39. Accordingly, we set the bandwidth $\lambda$ to 39 and use the global Moran's I index at this bandwidth for local model weight $\alpha$, which is 0.46.

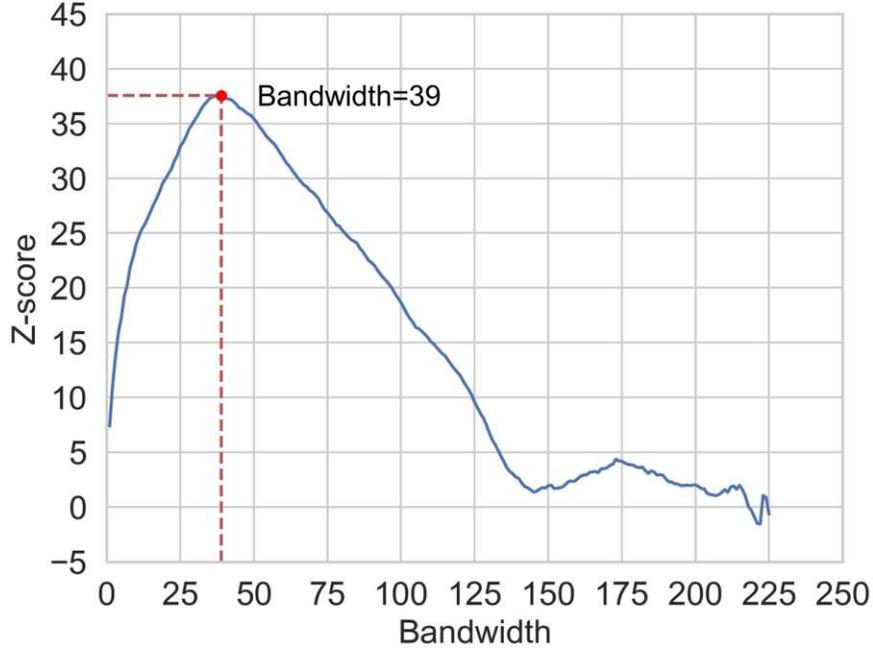

Figure 2. Incremental spatial autocorrelation plot for the example income dataset.

Table 1 summarizes the results of the three sets of experiments and the hyperparameters of the tested models. All performance scores are obtained via the same ten-fold cross-validation process. For the first set of experiments that compare PyGRF with RF, the PyGRF model achieves a higher $R^2$ than RF by 0.026 and a lower RMSE by 70.21. This result suggests that PyGRF improves over the RF model in terms of prediction accuracy, although the improvement is small. For the second set of experiments that evaluate the effectiveness of each improvement, we can see that the theory-informed hyperparameter determination (I1) slightly decreases the performance of PyGRF, while the local training sample expansion (I2) and spatially-weighted local prediction (I3) increase the model performance. If we compare the improvements of I2 and I3 with the improvement of PyGRF over RF, we can see that I2 and I3 bring in about 23.4% and 53.2% additional performance improvements respectively in terms of RMSE. While I1 slightly decreases the performance of the model compared with PyGRF, it substantially reduces the time needed for hyperparameter tuning (which will be discussed in the following paragraph). For the third set of experiments that examine the effectiveness of the proposed model improvements incrementally, we can see that the performance of the model increases gradually with the improvements added step-by-step. Adding all three model improvements (i.e., PyGRF + I1, I2, I3) achieves a higher performance than the default PyGRF model.

Table 1. A summary of results from the three sets of experiments on the example income dataset. I1, I2, and I3 represent the three model improvements respectively, which are: theory-informed hyperparameter determination (I1), local training sample expansion (I2), and spatially-weighted local prediction (I3).

| Models | ntree | mtry | $\lambda$ | $\alpha$ | $R^2$ | RMSE |
|---|---|---|---|---|---|---|
| *PyGRF* | 60 | 1 | 20 | 0.5 | 0.7212 | 1551.5797 |
| *Comparison with the RF model* | | | | | | |
| *RF* | 60 | 1 | / | / | 0.6954 | 1621.7899 |
| *Effectiveness of each improvement* | | | | | | |
| *PyGRF + I1* | 60 | 1 | 39 | 0.46 | 0.7191 | 1557.4078 |
| *PyGRF + I2* | 60 | 1 | 20 | 0.5 | 0.7271 | 1535.1649 |
| *PyGRF + I3* | 60 | 1 | 20 | 0.5 | **0.7345** | **1514.2185** |
| *Performance with each improvement added step-by-step* | | | | | | |
| *PyGRF + I1* | 60 | 1 | 39 | 0.46 | 0.7191 | 1557.4078 |
| *PyGRF + I1, I2* | 60 | 1 | 39 | 0.46 | 0.7210 | 1552.3255 |
| *PyGRF + I1, I2, I3* | 60 | 1 | 39 | 0.46 | **0.7231** | **1546.5187** |

Figure 3 shows the hyperparameter tuning times of the default PyGRF model and the models with the three improvements added step-by-step. All hyperparameter tuning is done using the same 70% data randomly selected from the original dataset. As can be seen, our proposed theory-informed hyperparameter determination (I1) substantially reduces the hyperparameter tuning time from over 28 minutes used by the PyGRF model to only about 1 minute, demonstrating an over 96% time saving. While our proposed local training sample expansion (I2) and spatially-weighted local prediction (I3) slightly increase model complexity, their increased hyperparameter tuning time is negligible. Note that all test models shown in Figure 3 are implemented in Python, and this experiment design ensures that the reduced hyperparameter tuning time indeed comes from the proposed model improvement rather than the use of a different programming language. In addition, the hyperparameter tuning time of the improved models includes the additional time of performing incremental spatial autocorrelation to ensure a fair comparison with the default trial-and-error approach.

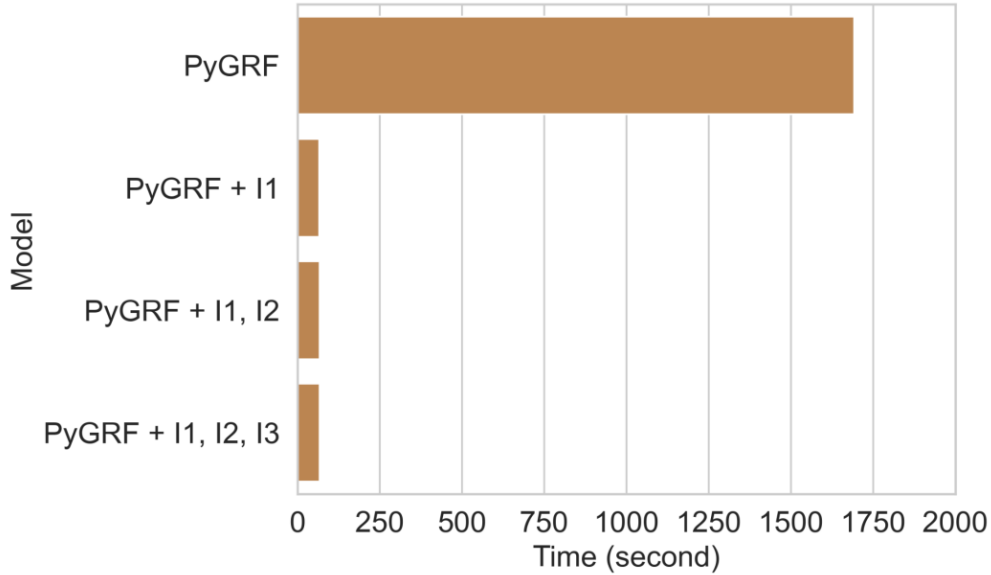

Figure 3. Hyperparameter tuning time of PyGRF and the three model improvements added step-by-step based on the example income dataset.

Overall, the experiment results suggest that the theory-informed hyperparameter determination (I1) can substantially reduce the time needed for finding suitable hyperparameters, and that local training sample expansion (I2) and spatially-weighted local prediction (I3) can increase model performance to some extent. The increases provided by I2 and I3 are fairly large compared with the increase of PyGRF over RF, but are small in terms of the absolute numbers in $R^2$ and RMSE. The limited performance increase of I3 is surprising given that it has the ability to reduce the influence of data outliers by leveraging multiple local RF models. We think that this is probably due to the lack of outliers in the experiment data. With curiosity, we simulate data outliers in the example income dataset by randomly replacing 1% of the data with large outlier values, and we then run the PyGRF model with and without I3. The result is shown in Figure 4. As can be seen, the simulated data outliers largely affect the default PyGRF model and result in largely-off local predictions and even a negative $R^2$ value. While the PyGRF with I3 is also affected by the outliers, it demonstrates a more robust performance with fewer large errors, more accurate local predictions, and a much higher $R^2$ value. This result suggests that I3 can increase the robustness of the PyGRF model by allowing the model to make fairly accurate predictions in the presence of data outliers.

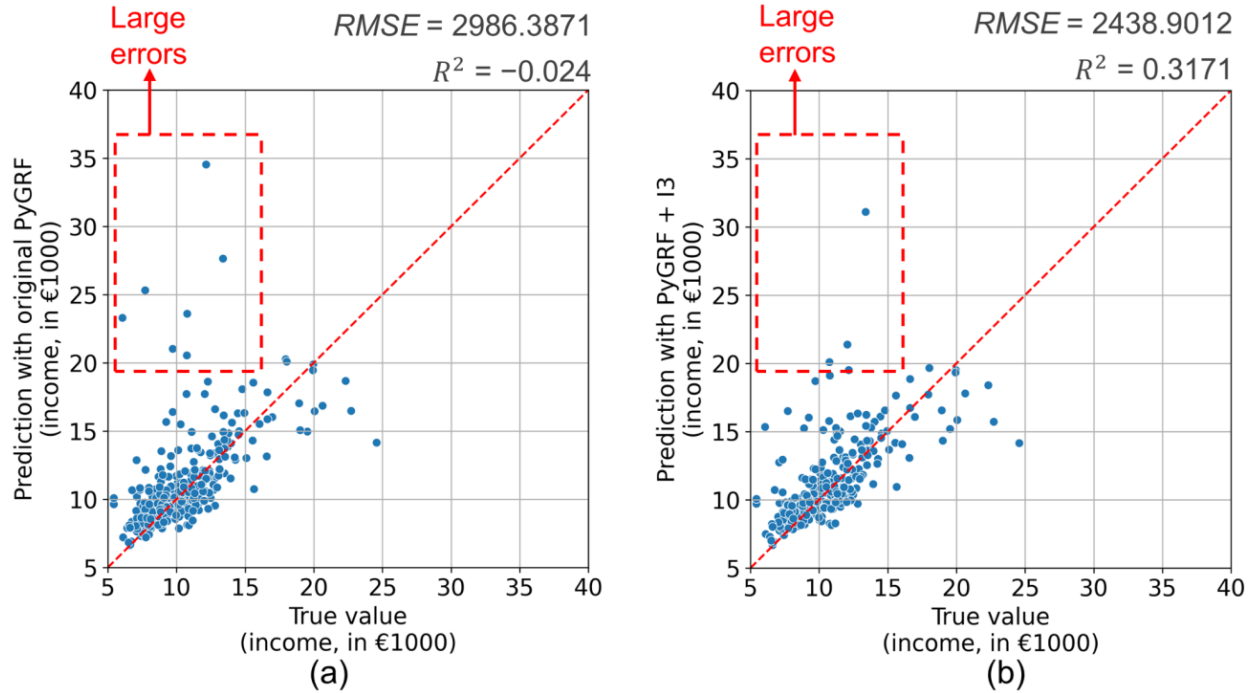

Figure 4. Experiment results based on the example income data with 1% simulated data outliers: (a) PyGRF model without I3; (b) PyGRF model with I3. (Note that the local weight is set to 1 to focus on local models, since I3 does not affect the global model. Other hyperparameters are: ntree: 60, mtry: 1, and bandwidth: 20).

## 5 Case studies in public health and natural disasters

In this section, we present two case studies in which we apply the developed PyGRF model and three improvements to real-world problems in the domains of public health and natural disasters. These two case studies serve the purposes of demonstrating the use of the model beyond the example dataset and further evaluating the performance of the proposed model improvements.

### 5.1 Obesity prevalence estimation in New York City

In the first case study, we use PyGRF to estimate neighborhood-level obesity prevalence in New York City (NYC). While obesity prevalence data can be obtained through surveys, model estimates are often necessary to fill in spatial and temporal data gaps (e.g., for obtaining data in geographic areas not covered by the survey or data in more recent years that are not available yet). Here, we assess the ability of PyGRF to estimate neighborhood-level obesity prevalence using socioeconomic and demographic variables. The dependent variable is obesity prevalence in NYC in 2018 obtained from the *PLACES* project of the Centers for Disease Control and Prevention (CDC). The geographic unit of analysis is census tracts. The independent variables include 21 socioeconomic and demographic factors organized in six categories: (1) race and ethnicity, (2) gender, marital status, and age, (3) education, (4) economic status, (5) housing conditions, and (6) urbanicity. The entire dataset was also used in our previous work (Zhou et al., 2022), and more details about the dependent and independent variables can be found in that article. We plot the neighborhood-level obesity prevalence in Figure 5(a).

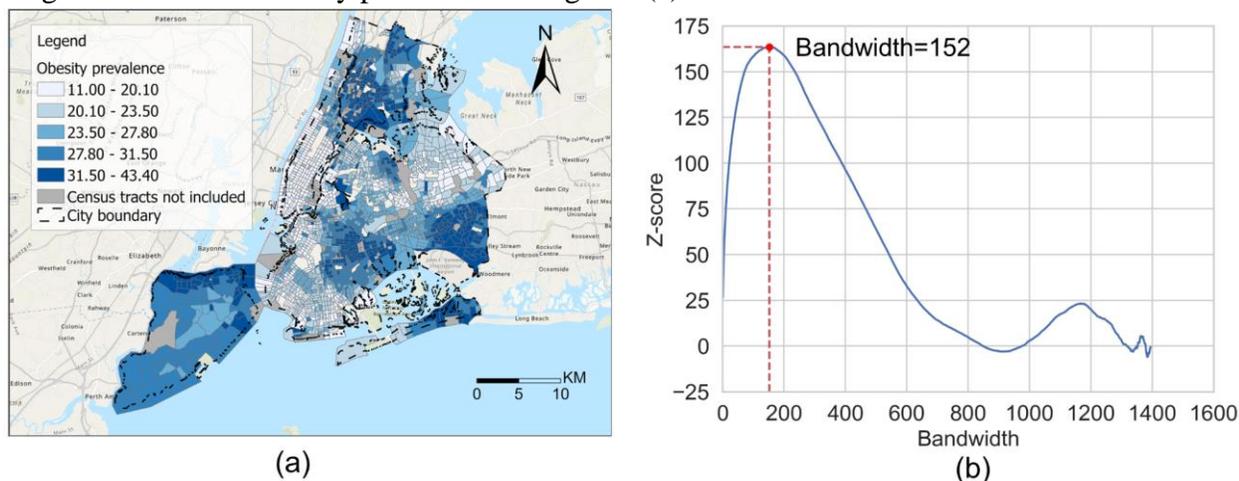

(a)   (b)

Figure 5. Neighborhood-level obesity prevalence in NYC and incremental spatial autocorrelation of the data: (a) a map visualization of obesity prevalence; (b) incremental spatial autocorrelation test result.

We conduct the same three sets of experiments, as done on the example income dataset previously, to compare PyGRF with RF, evaluate the effectiveness of each model improvement individually, and assess model performance change with each improvement added step-by-step. The search spaces of the four hyperparameters are set in a similar manner as done in the example income dataset: *ntree:* (0, $N/2$] with an interval of 100, where $N$ is the total number of samples in the training data (there are 1995 census tracts in total); *mtry:* $\{S, S/3, \sqrt{S}\}$, where $S$ is the number of independent variables; $\lambda$: [0.05 quantile, 0.95 quantile] of total samples with an interval of 100; $\alpha$: $\{0.25, 0.5, 0.75\}$. Again, PyGRF needs to tune all four hyperparameters, while the improved models will determine $\lambda$ and $\alpha$ based on the incremental spatial autocorrelation test. As shown in the test result in Figure 5(b), the obesity prevalence data of NYC has a different spatial autocorrelation pattern compared with the previous example income dataset. Based on the test result, we set the bandwidth $\lambda$ to 152 and local weight $\alpha$ to 0.4488 (given that the global Moran's I index at the bandwidth of 152 is 0.4488). All hyperparameter tuning is based on the same 70% data randomly selected from the entire dataset, and the performance of the models is measured via ten-fold cross-validation.

Table 2 shows the results from the three sets of experiments, and Figure 6 shows the time cost of hyperparameter tuning of the default PyGRF model and the models with improvements added step-by-step. For the first set of experiments, PyGRF achieves an increase of 0.037 in $R^2$ and a decrease of 0.368 in RMSE, compared with the RF model. For the second set of experiments, adding the theory-informed hyperparameter determination (I1) slightly decreases the performance of the model but substantially reduces the hyperparameter tuning time. As shown in Figure 6, the default PyGRF model uses more than 24 hours to find suitable values for the hyperparameters, while the theory-informed hyperparameter determination reduces the time to only about 3 hours while achieving a similar performance. Also shown in the result of the second set of experiments, the local training sample expansion (I2) and spatially-weighted local prediction (I3) both improve the performance of the PyGRF model with about 1.2% and 2.7% further improvement compared with the performance improvement of PyGRF over RF. For the third set of experiments, adding I1 and I2 achieves the best performance, while adding all three improvements leads to a slight decrease of model performance (a decrease of 0.0017 in $R^2$). Since the improvements brought by I2 and I3 are overall small (as shown in the second set of experiments), the slight performance decrease when all three improvements are added could be due to potential noise introduced when more local models are used for making predictions.

Table 2. A summary of results from the three sets of experiments on the obesity prevalence data of NYC. I1, I2, and I3 represent the three model improvements respectively, which are: theory-informed hyperparameter determination (I1), local training sample expansion (I2) , and spatially-weighted local prediction (I3).

| Models | ntree | mtry | $\lambda$ | $\alpha$ | $R^2$ | RMSE |
|---|---|---|---|---|---|---|
| *PyGRF* | 300 | S/3 | 149 | 0.75 | 0.9305 | 1.5517 |
| *Comparison with the RF model* | | | | | | |
| *RF* | 300 | S/3 | / | / | 0.8937 | 1.9192 |
| *Effectiveness of each improvement* | | | | | | |
| *PyGRF + I1* | 400 | S/3 | 152 | 0.4488 | 0.9241 | 1.6219 |
| *PyGRF + I2* | 300 | S/3 | 149 | 0.75 | 0.9309 | 1.5472 |
| *PyGRF + I3* | 300 | S/3 | 149 | 0.75 | **0.9314** | **1.5419** |
| *Performance with each improvement added step-by-step* | | | | | | |
| *PyGRF + I1* | 400 | S/3 | 152 | 0.4488 | 0.9241 | 1.6219 |
| *PyGRF + I1, I2* | 400 | S/3 | 152 | 0.4488 | **0.9246** | **1.6162** |
| *PyGRF + I1, I2, I3* | 400 | S/3 | 152 | 0.4488 | 0.9229 | 1.6341 |

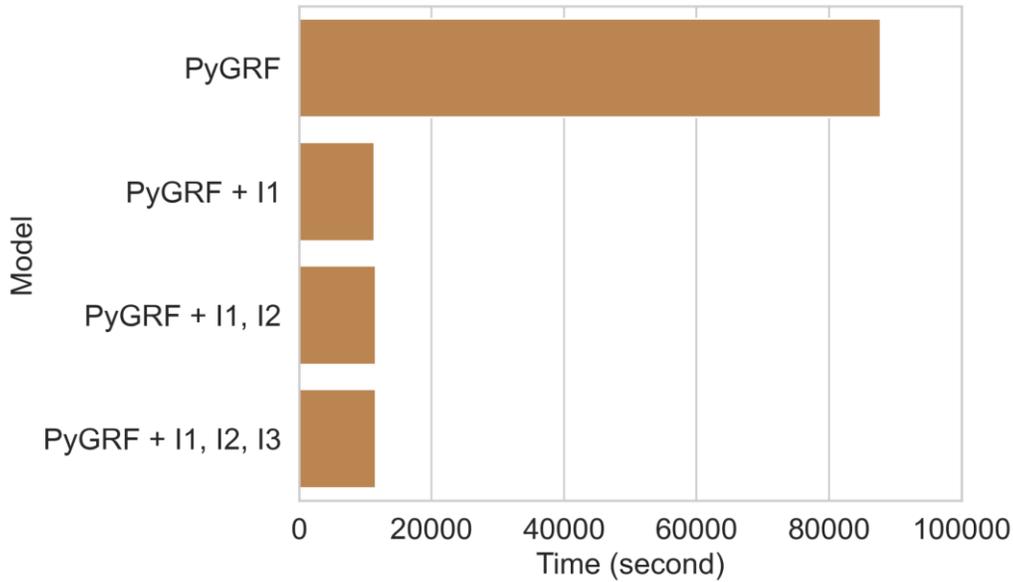

Figure 6. Hyperparameter tuning time of PyGRF and the three model improvements added step-by-step based on the obesity prevalence data of NYC.

One advantage of PyGRF is that it enables the exploration of both the global and local feature importance output by the global and local RF models contained in PyGRF. Here, we use the model, PyGRF + I1, I2, i.e., PyGRF with theory-informed hyperparameter determination and local training sample expansion, to explore feature importance. While the performance of PyGRF + I1, I2 is slightly lower than the default PyGRF (as shown in Table 2), it requires substantially less hyperparameter tuning time (only 13% of the time cost of PyGRF) and thus presents a more practical approach for model users. Figure 7 shows the global feature importance from the global RF models. Note that since ten-fold cross-validation is used, there are ten importance values for each feature in the box plot. As shown in the figure, two variables related to race and ethnicity show high importance for estimating neighborhood-level obesity prevalence in this case study, with *% Black*[2] ranked first and *% Asian*[3] in 2nd place. Three variables associated with housing condition, socioeconomic status, and poverty level are also ranked very high, with *median value units built*[4], *median household income*[5], and *% food stamp/SNAP*[6] ranked in the 3rd, 4th, and 5th places, respectively.

---

[2] Percentage of population in Black or African American

[3] Percentage of population in Asian

[4] Median value of the house units built (in dollars)

[5] Median household income

[6] Percentage of households received food stamp/supplemental nutrition assistance program (SNAP) in the past 12 months

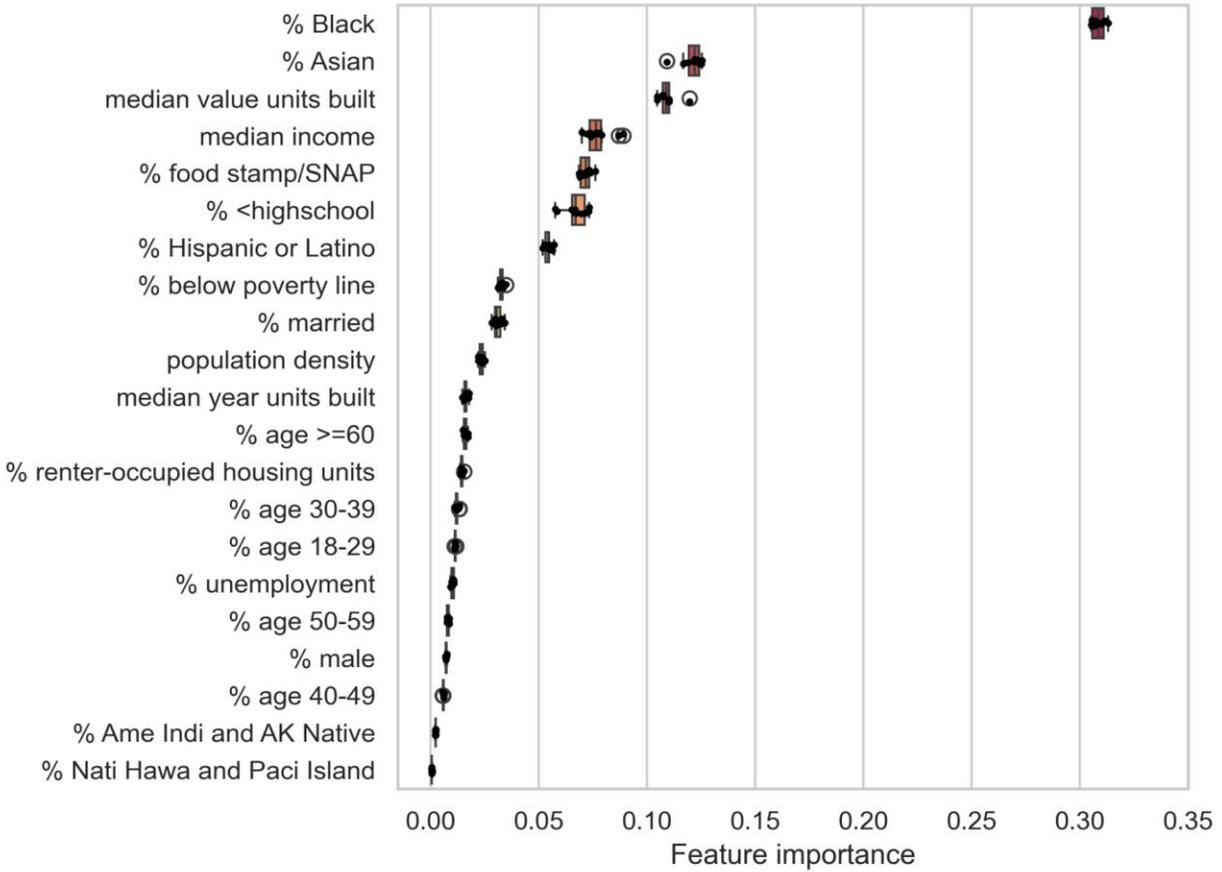

Figure 7. Box plot of the global feature importance for the case study of obesity prevalence estimation in NYC obtained from the PyGRF + I1, I2 model.

Next, we examine local feature importance and their spatial variation in the study area. In particular, we focus on two variables, namely *% Black* and *median household income*, shown to have high importance for obesity estimation based on the global model. Figure 8 shows the local feature importance of the two variables across NYC. Note that the value of feature importance is between 0 and 1. As can be seen, the feature importance of *% Black* varies largely across the city, from lower importance in most census tracts in middle Manhattan (with importance value in [0, 0.05)) to much higher importance in some peripheral areas of the city, e.g., the eastern side of Queens and the northern part of the Bronx (with importance value in [0.28, 0.45]). The local feature importance of *median household income* shows a largely different and almost reversed pattern: it seems to be more important for the local models to estimate neighborhood-level obesity prevalence in areas such as Brooklyn and the southern part of the Bronx, and seems to be less important in the eastern side of Queens. These spatial patterns can help researchers further develop hypotheses and examine the underlying reasons.

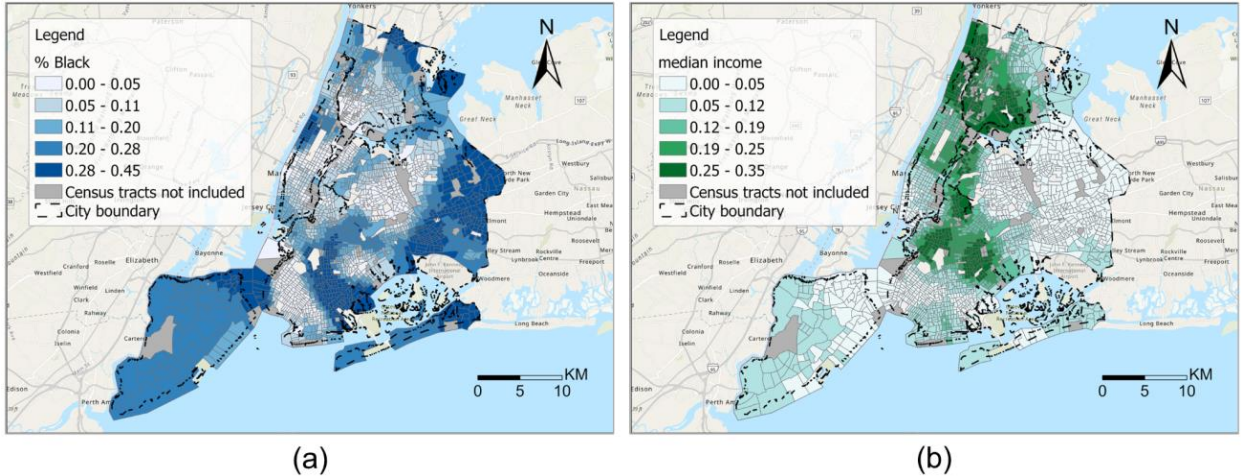

Figure 8. Map visualizations for the local feature importance of two variables: (a) *% Black*; (b) *median household income*.

## 5.2 Predicting help requests for winter storm preparation in Buffalo

In the second case study, we use PyGRF to predict potential help requests related to winter storms and blizzards in the city of Buffalo, USA. Buffalo experienced a severe blizzard in December 2022, during which many residents used the city's 311 call service to request help (Kaufman et al., 2023). The ability to predict potential help requests across different neighborhoods of the city can help emergency managers better prepare for future winter storms and blizzards. In this case study, the dependent variable is the number of 311-based help requests during the blizzard period obtained from the Open Data Portal of the City of Buffalo. The geographic unit for analysis is census block group (CBG), and the number of help requests is normalized by CBG population to obtain request count per person. Buffalo has a total of 290 CBGs, and the normalized 311 calls of the CBGs are shown in Figure 9(a). The independent variables include 18 factors that cover three aspects of each CBG: social vulnerability (Flanagan et al., 2011), physical vulnerability, and previous human behavior. These 18 independent variables are organized in six categories: (1) socioeconomic status, (2) household composition and disability, (3) minority status and language, (4) housing and transportation, (5) snow condition, and (6) historical 311 requests before the blizzard. The first four categories focus on social vulnerability, category (5) focuses on physical vulnerability, and category (6) focuses on previous human behavior. Details of these independent variables are provided in Supplementary Table S1.

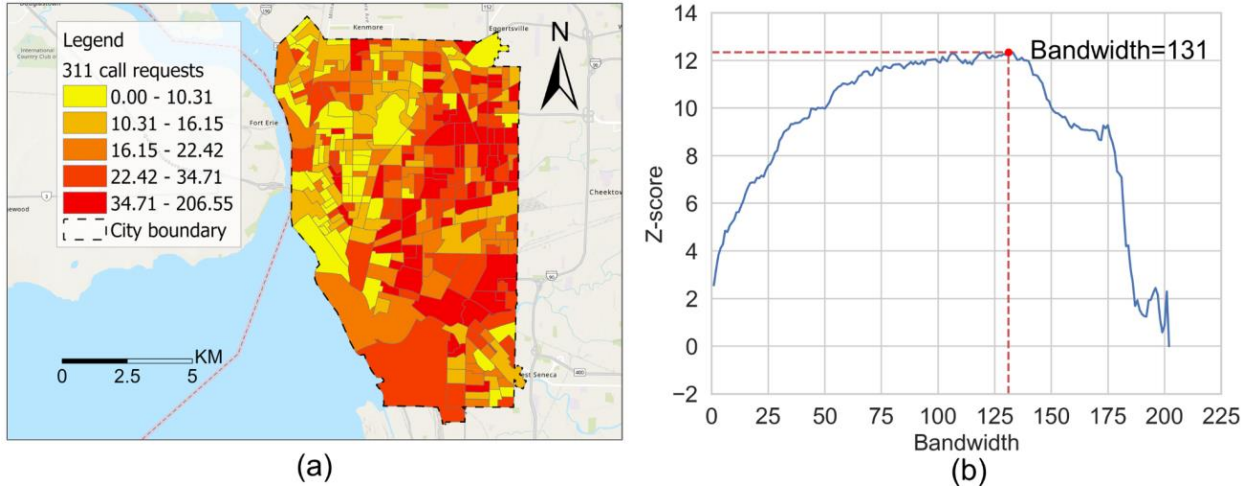

Figure 9. Help requests during 2022 Buffalo blizzard and incremental spatial autocorrelation of the data: (a) CBG-level normalized help requests from 12/19/2022 to 1/1/2023; (b) incremental spatial autocorrelation test result.

We conduct the same three sets of experiments in this case study to compare PyGRF with RF, evaluate the effectiveness of each model improvement individually, and assess performance change with each improvement added step-by-step. The hyperparameters are tuned in a similar way as in the first case study. We tune four hyperparameters for the default PyGRF model, and determine the bandwidth $\lambda$ and local model weight $\alpha$ based on the incremental spatial autocorrelation test shown in Figure 9(b). The bandwidth $\lambda$ is set to 131 and local weight $\alpha$ is set to 0.0444 (a weak but significant spatial autocorrelation is observed). Table 3 shows the results from the three sets of experiments, and Figure 10 shows the time of hyperparameter tuning of four models. For the first set of experiments, PyGRF improves over the RF model, with an increase of 0.0298 in $R^2$ and a decrease of 0.384 in RMSE. For the second set of experiments, the theory-informed hyperparameter determination (I1) achieves not only a substantial decrease of hyperparameter tuning time (shown in Figure 10) but also an increase of model performance. The local training sample expansion (I2) is not activated since there already exist sufficient local samples for training the RF model (the number of local training samples is larger than two times of the tree number). The spatially-weighted local prediction (I3) also improves the performance of the model, demonstrating a further 94.2% improvement compared with the improvement of PyGRF over RF. For the third set of experiments, adding all three model improvements achieves better performance than the default *PyGRF* model while using only a small fraction of the time for hyperparameter tuning (about 2 minutes compared with the 32 minutes used by the default PyGRF model).

Table 3. A summary of results from the three sets of experiments on the 311 help request data of Buffalo. I1, I2, and I3 represent the three model improvements respectively, which are: theory-informed hyperparameter determination (I1), local training sample expansion (I2), and spatially-weighted local prediction (I3).

| Models | ntree | mtry | $\lambda$ | $\alpha$ | $R^2$ | RMSE |
|---|---|---|---|---|---|---|
| *PyGRF* | 20 | $\sqrt{S}$ | 125 | 0.75 | 0.3898 | 15.8898 |
| *Comparison with the RF model* | | | | | | |
| *RF* | 20 | $\sqrt{S}$ | / | / | 0.3600 | 16.2740 |
| *Effectiveness of each improvement* | | | | | | |
| *PyGRF + I1* | 60 | S/3 | 131 | 0.0444 | **0.4205** | **15.4860** |
| *PyGRF + I2 \** | 20 | $\sqrt{S}$ | 125 | 0.75 | 0.3898 | 15.8898 |
| *PyGRF + I3* | 20 | $\sqrt{S}$ | 125 | 0.75 | 0.4173 | 15.5280 |
| *Performance with each improvement added step-by-step* | | | | | | |
| *PyGRF + I1* | 60 | S/3 | 131 | 0.0444 | 0.4205 | 15.4860 |
| *PyGRF + I1, I2 \** | 60 | S/3 | 131 | 0.0444 | 0.4205 | 15.4860 |
| *PyGRF + I1, I2, I3* | 60 | S/3 | 131 | 0.0444 | **0.4205** | **15.4854** |

\*The strategy of local training sample expansion (I2) is not activated by the model, since there already exist sufficient local training samples.

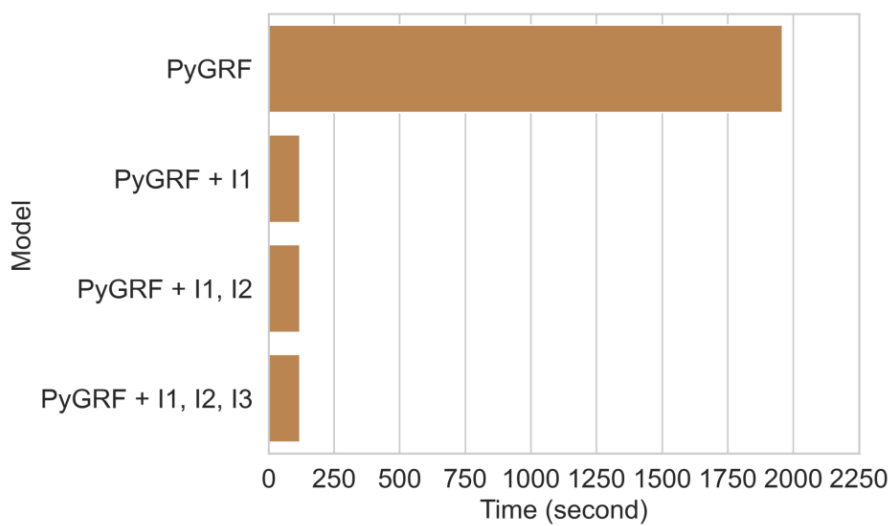

Figure 10. Hyperparameter tuning time of PyGRF and the three model improvements added step-by-step based on the 311 help request data of Buffalo.

We further explore the global and local feature importance output by the PyGRF + I1, I2, I3 model. Figure 11 shows the global feature importance based on the ten-fold cross-validation result. As can be seen, two variables related to previous 311 call behavior hold very high importance, ranking as the 1st and 2nd. The *% minority* and *% 65 older* are ranked as 3rd and 5th respectively. The variable related to snow depth is ranked as the 4th. The other variables exhibit relatively lower importance for predicting 311 help requests in this case study.

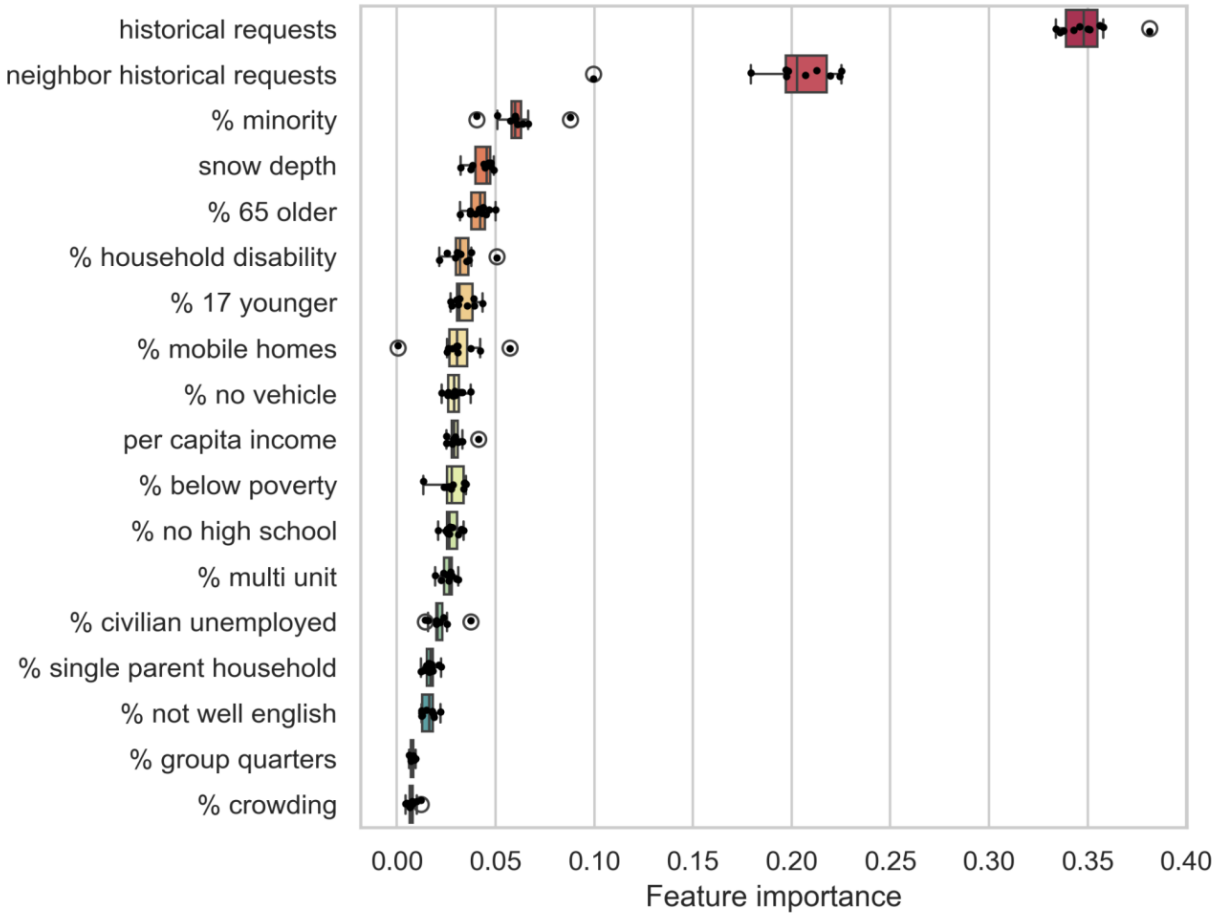

Figure 11. Box plot of the global feature importance for the case study of 311 help request prediction in Buffalo obtained from the PyGRF + I1, I2, I3 model.

Figure 12 shows the local feature importance of two variables, *historical requests* and *% minority*, across the study area. As can be seen, *historical requests* seem to be more important for local models to predict 311 help requests in the western areas of Buffalo (which is the core city region) but relatively less important in the northern areas. This result seems to suggest that the number of 311 calls in the core city region of Buffalo is largely affected by the extent to which the residents have previously used the 311 call service. Meanwhile, the local feature importance of *% minority* shows a different spatial pattern, with higher importance in the central and northern areas of the city but lower importance in the southern areas. These spatial patterns can be further investigated to identify the underlying reasons.

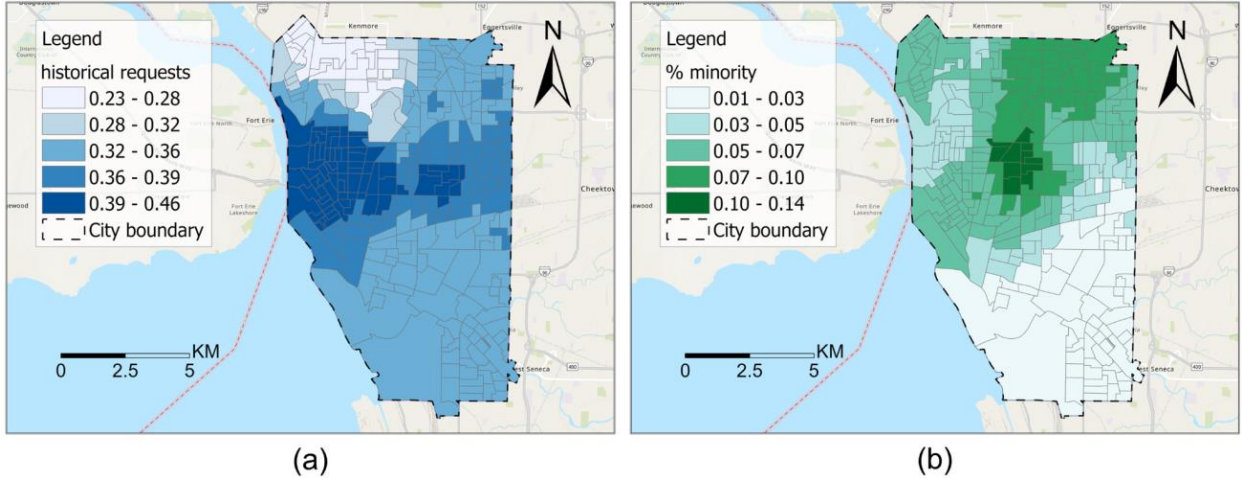

Figure 12. Map visualizations for the local feature importance of two variables in the case study of 311 help requests prediction in Buffalo obtained from the PyGRF + I1, I2, I3 model: (a) *historical requests*; (b) *% minority*.

## 6 Discussion

### 6.1 Effectiveness of the proposed model improvements

We have proposed three model improvements for GRF building on the work of Georganos et al. (2021; 2022). The three improvements are: theory-informed hyperparameter determination (I1), local training sample expansion (I2), and spatially-weighted local prediction (I3). The results from the example data and two case studies suggest that the theory-informed hyperparameter determination is highly effective in reducing the time cost of hyperparameter tuning. By first understanding the spatial autocorrelation of the data, rather than directly training and testing many GRF models based on different hyperparameter combinations, we reduce 96%, 87%, and 94% of hyperparameter tuning time for the example income dataset, NYC obesity prevalence dataset, and Buffalo help request dataset respectively. This time cost reduction is important, as it enables researchers and machine learning practitioners to explore the use of GRF in their data more efficiently. For example, in the case study of obesity prevalence estimation in NYC, the dataset has a moderate size of 1995 data records; yet, it takes over one day for the trial-and-error approach to find suitable hyperparameter values. Our theory-informed hyperparameter determination reduces this time cost to about 3 hours, and makes it more realistic for model users to explore GRF. In terms of model performance, I1 slightly reduces prediction accuracy on the example income dataset and the NYC obesity prevalence dataset, while increasing accuracy on the Buffalo help request dataset. It is worth noting that the improvement is compared against optimized GRF models in all experiments. In situations when there is not enough time to perform comprehensive hyperparameter tuning, GRF can have much lower performance based on arbitrarily selected hyperparameter values, as shown in the original paper (Georganos et al., 2021). Thus, our theory-informed hyperparameter determination allows the GRF model to still achieve a close-to-optimal performance when there is only limited time for hyperparameter tuning.

For local training sample expansion (I2) and spatially-weighted local prediction (I3), our experiments suggest that they are overall effective in improving model performance. In particular, our second set of experiments that evaluate the effectiveness of each improvement individually show that the improvements of I2 and I3 can be substantial compared with the improvement brought by GRF to RF. Nevertheless, the absolute numbers of improvements in $R^2$ and RMSE are small. For I2, although it increases the size of the local training samples, it does not bring in new information for the local RF models. In our earlier experiments, we also tried another approach to address the issue of insufficient local training samples by reducing the complexity of local RF models (e.g., reducing the number of trees in local RF models). However, we found that this approach did not work as well as our current I2, i.e., expanding local training samples while keeping the same number of trees in local RF models. The issue of lacking sufficient local training data might be better addressed when additional local data are made available. For I3, our experiments with simulated data outliers in Section 4.3 have shown that it can increase model robustness when outliers are present in the data. In our developed PyGRF package, we have incorporated all three improvements. To provide more flexibility for model users, we have implemented PyGRF in a way that allows the user to turn on and off any of the three model improvements. Thus, the users of PyGRF can choose to use the original version of GRF by turning off all improvements, but can also turn on any combination or all improvements to further increase model performance.

## 6.2 Limitations

This research is not without limitations. First, while we have tested PyGRF and the proposed model improvements on three different datasets, testing and using the model on more datasets in a variety of domains can help further understand its advantages and limitations. To this end, we hope that our implemented PyGRF package published in the widely used *pip* package management system, source code shared on GitHub, and the Jupyter Notebooks of the case studies can help more machine learning and GeoAI practitioners to try this model and further improve it. Second, while PyGRF improves over GRF and RF, it is only one of the possible models that we can choose from, and it may not always provide the best performance among all possible models for a given dataset. In practice, it is necessary to test multiple models and choose the most suitable one (Wiedemann et al., 2023). Third, our current PyGRF model focuses on regression tasks, and this focus is consistent with the original R-based GRF model. However, random forest has also been used for classification tasks. Future studies could extend PyGRF for handling classification tasks, and may assess its classification accuracy and the usefulness of local feature importance derived from those tasks.

## 7 Conclusions

In this work, we have proposed three model improvements, including theory-informed hyperparameter determination, local training sample expansion, and spatially-weighted local prediction, to address limitations identified from the current GRF model. We have also developed

PyGRF, which is a Python-based Geographical Random Forest model and package, to incorporate these improvements. We have evaluated the consistency between the PyGRF model and the original R-based model, and have applied PyGRF and the improved models to two case studies in public health and natural disasters. The results show that the PyGRF provides overall consistent output with the R-based GRF model, and the three proposed improvements increase model performance while substantially reducing the time cost for hyperparameter tuning. This work contributes to both GeoAI methods and the development of open-source GIS packages. Regarding the latter, we have published PyGRF on the Python package management system pip, shared its source code on GitHub, and provided Jupyter Notebooks for the two case studies. We hope that this effort can make it easier for others to use this open-source package. While PyGRF is not without limitations, we hope that it can serve as a tool for researchers and machine learning practitioners to explore spatial variation of local feature importance, improve prediction accuracy, and derive more insights from geospatial data.

## Acknowledgments

This work is supported by the U.S. National Science Foundation under Grant Nos. BCS-2117771 and BCS-2416886. Any opinions, findings, and conclusions or recommendations expressed in this material are those of the authors and do not necessarily reflect the views of the National Science Foundation.

## Data Availability Statement

The data that support the findings of this study are available on GitHub at https://github.com/geoai-lab/PyGRF.

## References

Aguirre-Gutiérrez, J., Rifai, S., Shenkin, A., Oliveras, I., Bentley, L. P., Svátek, M., Girardin, C. A., Both, S., Riutta, T., & Berenguer, E. (2021). Pantropical modelling of canopy functional traits using Sentinel-2 remote sensing data. *Remote Sensing of Environment*, *252*, 112122.

Anselin, L. (2009). Spatial regression. *The SAGE Handbook of Spatial Analysis*, *1*, 255–276.

Breiman, L. (2001). Random forests. *Machine Learning*, *45*, 5–32.

Brunsdon, C., Fotheringham, S., & Charlton, M. (1998). Geographically weighted regression. *Journal of the Royal Statistical Society: Series D (The Statistician)*, *47*(3), 431–443.

Chang, T., Hu, Y., Taylor, D., & Quigley, B. M. (2022). The role of alcohol outlet visits derived from mobile phone location data in enhancing domestic violence prediction at the neighborhood level. *Health & Place*, *73*, 102736.

https://doi.org/10.1016/j.healthplace.2021.102736

DeAngelis, D. L., & Yurek, S. (2017). Spatially explicit modeling in ecology: A review. *Ecosystems*, *20*(2), 284–300.

Feng, L., Wang, Y., Zhang, Z., & Du, Q. (2021). Geographically and temporally weighted neural network for winter wheat yield prediction. *Remote Sensing of Environment*, *262*, 112514.

Flanagan, B. E., Gregory, E. W., Hallisey, E. J., Heitgerd, J. L., & Lewis, B. (2011). A Social Vulnerability Index for Disaster Management. *Journal of Homeland Security and Emergency Management*, *8*(1). https://doi.org/10.2202/1547-7355.1792

Fotheringham, A. S., Brunsdon, C., & Charlton, M. (2003). *Geographically weighted regression: The analysis of spatially varying relationships*. John Wiley & Sons.

Fotheringham, A. S., Yang, W., & Kang, W. (2017). Multiscale geographically weighted regression (MGWR). *Annals of the American Association of Geographers*, *107*(6), 1247–1265.

Gao, S., Li, M., Liang, Y., Marks, J., Kang, Y., & Li, M. (2019). Predicting the spatiotemporal legality of on-street parking using open data and machine learning. *Annals of GIS*, *25*(4), 299–312.

Georganos, S., Grippa, T., Niang Gadiaga, A., Linard, C., Lennert, M., Vanhuysse, S., Mboga, N., Wolff, E., & Kalogirou, S. (2021). Geographical random forests: A spatial extension of the random forest algorithm to address spatial heterogeneity in remote sensing and population modelling. *Geocarto International*, *36*(2), 121–136.

Georganos, S., & Kalogirou, S. (2022). A forest of forests: A spatially weighted and computationally efficient formulation of geographical random forests. *ISPRS International Journal of Geo-Information*, *11*(9), 471.

Goodchild, M. (2001). Issues in spatially explicit modeling. *Agent-Based Models of Land-Use and Land-Cover Change*, 13–17.

Grekousis, G., Feng, Z., Marakakis, I., Lu, Y., & Wang, R. (2022). Ranking the importance of demographic, socioeconomic, and underlying health factors on US COVID-19 deaths: A geographical random forest approach. *Health & Place*, *74*, 102744.

Griffith, D. A. (2003). *Spatial autocorrelation and spatial filtering: Gaining understanding through theory and scientific visualization*. Springer-Verlag.

Gupta, J., Molnar, C., Xie, Y., Knight, J., & Shekhar, S. (2021). Spatial Variability Aware Deep Neural Networks (SVANN): A General Approach. *ACM Transactions on Intelligent Systems and Technology (TIST)*, *12*(6), 1–21.

Hagenauer, J., & Helbich, M. (2022). A geographically weighted artificial neural network. *International Journal of Geographical Information Science*, *36*(2), 215–235.

Ho, T. K. (1995). Random decision forests. *Proceedings of 3rd International Conference on Document Analysis and Recognition*, *1*, 278–282.

Hu, Y., Goodchild, M., Zhu, A.-X., Yuan, M., Aydin, O., Bhaduri, B., Gao, S., Li, W., Lunga, D., & Newsam, S. (2024). A five-year milestone: Reflections on advances and limitations in GeoAI research. *Annals of GIS*, 1–14.

Hu, Y., Quigley, B., & Taylor, D. (2021). Human mobility data and machine learning reveal geographic differences in alcohol sales and alcohol outlet visits across U.S. states during COVID-19. *PLOS ONE*, *16*(12), e0255757.

Islam, M. D., Li, B., Lee, C., & Wang, X. (2021). Incorporating spatial information in machine learning: The Moran eigenvector spatial filter approach. *Transactions in GIS*.

Janowicz, K., Gao, S., McKenzie, G., Hu, Y., & Bhaduri, B. (2020). *GeoAI: spatially explicit artificial intelligence techniques for geographic knowledge discovery and beyond*. *34*(4), 625–636.

Kaufman, S. M., Zimmerman, R., Ozbay, K., Smith, A., Lambson, S. H., Curry, C., Jeng, E., Gao, J., & Kaval, E. (2023). *Lessons Learned from the Buffalo Blizzard: Recommendations for Strengthening Preparedness and Recovery Efforts*.

Li, L. (2019). Geographically weighted machine learning and downscaling for high-resolution spatiotemporal estimations of wind speed. *Remote Sensing*, *11*(11), 1378.

Li, Z. (2022). Extracting spatial effects from machine learning model using local interpretation method: An example of SHAP and XGBoost. *Computers, Environment and Urban Systems*, *96*, 101845.

Mai, G., Hu, Y., Gao, S., Cai, L., Martins, B., Scholz, J., Gao, J., & Janowicz, K. (2022). Symbolic and subsymbolic GeoAI: Geospatial knowledge graphs and spatially explicit machine learning. *Trans GIS*, *26*(8), 3118–3124.

Masrur, A., Yu, M., Mitra, P., Peuquet, D., & Taylor, A. (2022). Interpretable machine learning for analysing heterogeneous drivers of geographic events in space-time. *International Journal of Geographical Information Science*, *36*(4), 692–719.

O'Sullivan, D., Gahegan, M., Exeter, D. J., & Adams, B. (2020). Spatially explicit models for exploring COVID-19 lockdown strategies. *Transactions in GIS*, *24*(4), 967–1000.

Quevedo, R. P., Maciel, D. A., Uehara, T. D. T., Vojtek, M., Renno, C. D., Pradhan, B., Vojtekova, J., & Pham, Q. B. (2022). Consideration of spatial heterogeneity in landslide susceptibility mapping using geographical random forest model. *Geocarto International*, *37*(25), 8190–8213.

Tobler, W. R. (1970). A Computer Movie Simulating Urban Growth in the Detroit Region. *Economic Geography*, *46*, 234–240. https://doi.org/10.2307/143141

Wiedemann, N., Martin, H., & Westerholt, R. (2023). Benchmarking Regression Models Under Spatial Heterogeneity. *12th International Conference on Geographic Information Science (GIScience 2023)*.

Yan, B., Janowicz, K., Mai, G., & Gao, S. (2017). From ITDL to Place2Vec–Reasoning About Place Type Similarity and Relatedness by Learning Embeddings From Augmented Spatial Contexts. *Proceedings of SIGSPATIAL*, *35*, 1–10.

Zhou, R. Z., Hu, Y., Tirabassi, J. N., Ma, Y., & Xu, Z. (2022). Deriving neighborhood-level diet and physical activity measurements from anonymized mobile phone location data for enhancing obesity estimation. *International Journal of Health Geographics*, *21*(1), 1–18.